\documentclass{PoS}
\usepackage{cite, caption, subcaption}

\title{The ILC Potential for Discovering New Particles}

\ShortTitle{Discovering New Particles at ILC}

\author{\speaker{Jenny List}\thanks{On behalf of the LCC Physics Working Group}\\
        DESY\\
        E-mail: \email{jenny.list@desy.de}}

\abstract{The LHC did not discover new particles beyond the Standard Model Higgs boson at 7 and 8\,TeV, or in the first data samples at 13\,TeV.   However, the complementary nature of physics with $e^+e^-$ collisions still offers many interesting scenarios in which new particles can be discovered at the ILC.  These scenarios take advantage of the capability of $e^+e^-$ collisions to observe particles with missing energy and small mass differences, to observe mono-photon events with precisely controled backgrounds, and to observe the full range of exotic decay modes of the Higgs boson. The searches that an $e^+e^-$ collider makes possible are particularly important for models of dark matter involving a dark sector with particles above the modest energy reach of fixed-target experiments. In this talk, we will review the opportunities that the ILC offers for new particle discovery.}

\FullConference{The European Physical Society Conference on High Energy Physics\\
		5-12 July, 2017\\
		Venice}

\begin{document}

\section{Introduction}
The International Linear Collider (ILC) is a proposed electron-positron collider with a center-of-mass energy tunable between $200$ and $500$\,GeV, and upgradable to $1$\,TeV. The design luminosity at $\sqrt{s}=500$\,GeV is $1.8 \cdot 10^{34}$\,cm$^{-2}$s$^{-1}$, which can be doubled in a luminosity upgrade. The electron and positron beams are forseen to be polarised to $\pm80\%$ and $\pm30\%$, respectively. The Technical Design Report of the ILC was published in 2013~\cite{Behnke:2013xla,Baer:2013cma,Adolphsen:2013jya,Adolphsen:2013kya,Behnke:2013lya}, and the project  is currently being considered by the Japanese government.
This presentation at the EPS-HEP 2017 conference is based on a recent document~\cite{Fujii:2017ekh} by the LCC Physics Working Group, which contains more information on all the points summarized here.

\section{New Properties of the Higgs Boson and the Top Quark}
\begin{figure}[b]
\begin{center}
   \begin{subfigure}{.475\hsize}
      \includegraphics[width=\textwidth]{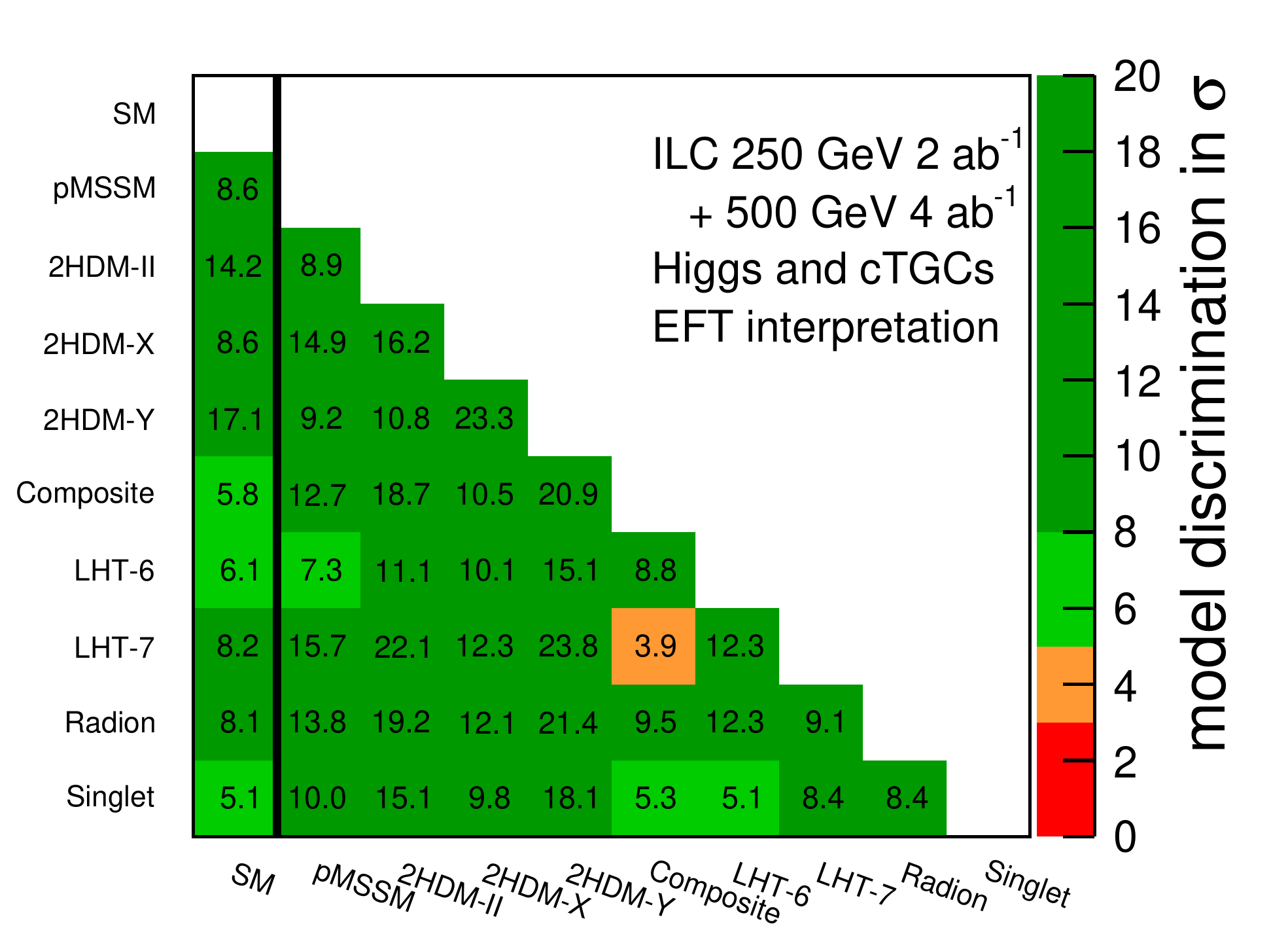}
         \subcaption{} \label{fig:matrix}
   \end{subfigure} 
   \hspace{0.03\hsize} 
   \begin{subfigure}{.475\hsize}
      \includegraphics[width=\textwidth]{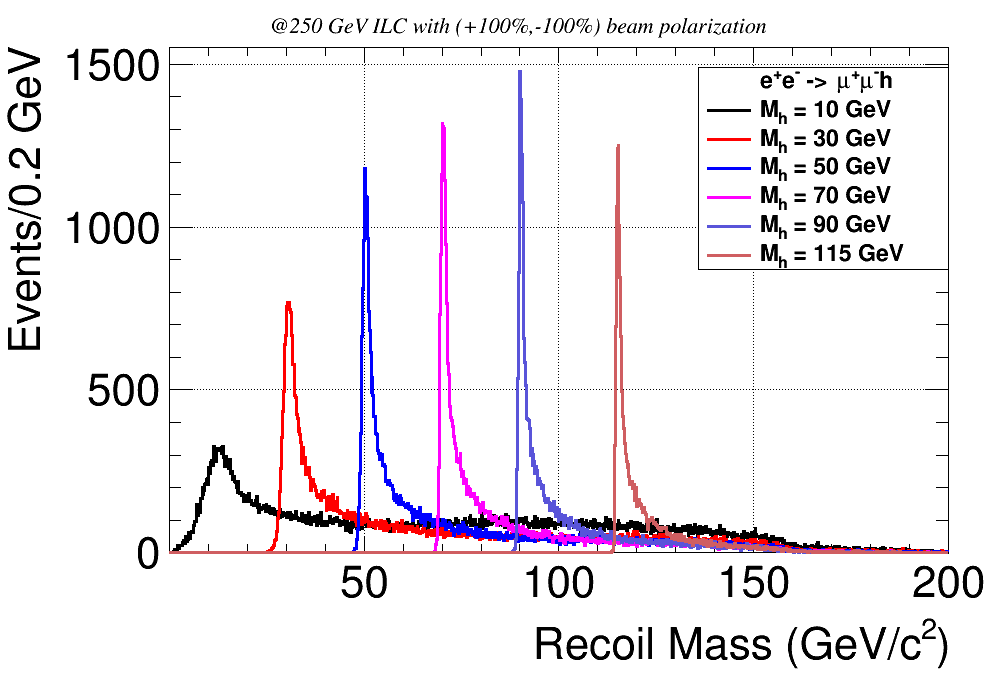}
         \subcaption{} \label{fig:recoil}
   \end{subfigure}  
\end{center}
\caption{(a) Significance for distinguishing various BSM benchmark points by ILC precision measurements. For exact definitions of the model points see~\cite{Barklow:2017suo}. (b) Generator-level recoil mass spectrum for various candidate Higgs masses, arbitrary normalisation.}
\label{fig:fig1}
\end{figure}

The most widely known way to discover new physics at an $e^+e^-$ collider 
is via precision measurements, in particular of the Higgs boson and the top quark,
but also of the $W$ and $Z$ bosons. While the precisions achievable at the ILC
on their couplings have been thoroughly evaluated --- including estimates of the systematic uncertainties --- already a few years ago~\cite{Fujii:2015jha}, we presented
at this conference for the first time how a joint interpretation of Higgs and electroweak precision measurements in the framework of effective field theory can serve to distinguish
various new physics benchmarks a) from the SM and b) among each other~\cite{Barklow:2017suo}, c.f.\ Fig.~\ref{fig:matrix}.

\section{Additional Higgs Bosons}
The recoil technique which uses the known initial state of a lepton collider to 
reconstruct a particle produced in association with a $Z$ boson independently of its
decay modes is well-established in the case of the 125-GeV Higgs boson~\cite{Yan:2016xyx}.
However it can also be applied to probe for additional (Higgs) bosons at lower masses, as
illustrated by the generator-level recoil mass distributions in Fig.~\ref{fig:recoil}.
Although this mass region has been probed already by LEP, the much higher luminosity of the ILC, in combination with its polarised beams, will allow to probe much smaller couplings
to the $Z$ boson. A full-simulation study of this channel is currently ongoing within the ILD concept group.

\section{Supersymmetry}
The supersymmetric partners of the Higgs boson(s), the Higgsinos, are well motived to
be light by naturalness arguments, but also to have small (few GeV) or even very small
(sub-GeV) mass splittings. Both cases offer very interesting discovery potential at the ILC. In the case of sub-GeV mass splittings, ILC measurements can probe values of
the Bino and Wino mass parameters in the 10-TeV range~\cite{Berggren:2013vfa}. More recently, the case of somewhat larger mass differences has been studied in full detector simulation. Permille-level mass and percent-level cross-section measurements can be performed, allowing to a) predict the rest of the SUSY mass spectrum, b) test the unification of gaugino masses and determine its energy scale and c) distinguish between different underlying GUT-scale models, c.f.~\cite{Baer:2016new} and the contribution to this conference by S.~Lehtinen.

Another unique asset of lepton colliders is the possibility to perform practically loop-hole free searches for $R$-parity conserving supersymmetry by systematically covering  pair production of the next-to-lightest-SUSY particle, which can only decay into its SM partner and the lightest SUSY particle. This is very similar to the spirit of simplified-model searches popular at the LHC, with the important difference that --- unlike at hadron colliders --- no strong model assumptions need to be made~\cite{Berggren:2013vna}. 

\section{Dark Matter}
The ILC will also make important contributions to the exploration of dark sectors with particles from just above the mass range of fixed target experiments, which currently probe up to a few 100\,MeV and might reach 1\,GeV e.g.\ with SHiPS~\cite{Dobrich:2015jyk}, up to half the center-of-mass energy for pair production, and much higher than that for mediator particles. 

\begin{figure}
\begin{center}
      \includegraphics[width=0.7\textwidth]{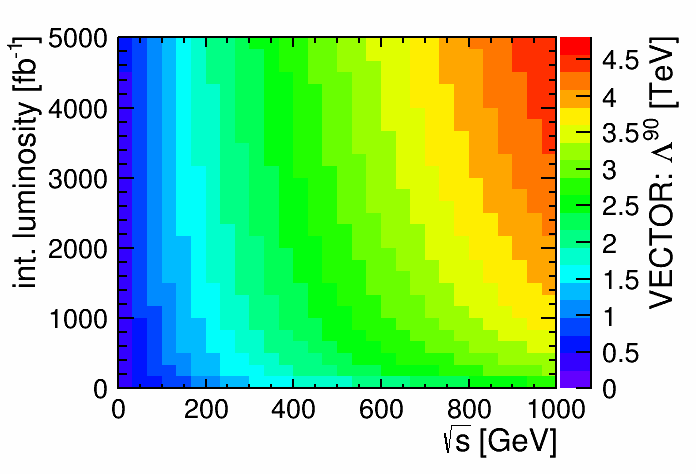}
\end{center}
\caption{New physics scales $\Lambda$ probed by the ILC at various center-of-mass energies and integrated luminosities for a WIMP with a mass of $10$\,GeV coupling through a vector-like operator to electrons~\cite{Habermehl:2017dxh}.}
\label{fig:WIMP}
\end{figure}

A famous example is the search for WIMPs in the mono-photon channel, in analogy to mono-X searches at hadron colliders, but probing directly the couplings of WIMPs to electrons. Due to their clean environment, they offer sensitivity to much smaller couplings or much higher energy scales of new physics\footnote{Note that at lepton colliders, the probed new physics scale $\Lambda$ is typically much larger than the center-of-mass energy, therefore the EFT approach does not suffer from the same problems limiting its applicability at the LHC.}. In both aspects, lepton collider searches are thus complementary to direct detection and searches at hadron colliders. Figure~\ref{fig:WIMP}
shows the energy scales probed at the ILC as a function of the center-of-mass energy and  the integrated luminosity~\cite{Habermehl:2017dxh}. E.g.\ for the nominal integrated luminosity of $4$\,ab$^{-1}$ at 500\,GeV, new physics scales of up to 3\,TeV can be probed. Note that beam polarisation plays an essential role here to reduce the SM neutrino background by nearly 2 orders of magnitude.

If the Dark Matter particle is part of a richer spectrum, then the precision determination
of masses and cross sections at the ILC allows to extract the parameters of the underlying
model and to determine the relic density as predicted by this model for comparison
with cosmological observations. For a recent example c.f.~\cite{Lehtinen:2016qis}.

\section{Neutrinos}
Some of the possible mechanisms to explain the smallness of neutrino masses and their mixing pattern are observable at colliders. For instance in SUSY models with bilinear $R$-parity violation, the branching ratios of the neutralinos into the various lepton flavours are related to the neutrino mixing angles. Whereas the lack of missing transverse energy makes RPV signatures more difficult to discover at the LHC, this is not the case at a lepton collider. Therefore neutralino pair production would be easily discoverable, even in the initial run of the ILC, for selectron masses of up to $2$\,TeV~\cite{List:2013dga}.
The measurement of the neutralino branching ratios then allows to derive a prediction
for the athmospheric mixing angle with a precision comparable to the current measurements
from oscillation data. Agreement or disagreement of the two values would confirm or disprove bilinear RPV SUSY as mechanism for neutrino mass generation.

\section{Conclusions}
In addition to its bread-and-butter precision program, the ILC offers significant potential to discover well-motivated new particles. Several examples were given in
this presentation. All rely on the well-appreciated properties of lepton colliders,
namely the well-defined initial state, the clean environment and the electroweak production rates, which together allow a trigger-less operation of the detectors, as
well as a nearly democratic production of all particles with electroweak charges.
But they also profit critically from the special assets of linear colliders, in particular the extendability in energy and the beam polarisation. In all the cases presented
here, the ILC's potential complements perfectly the discovery reach of the LHC.

\section*{Acknowledgements}
The LCC Physics Working Group thanks the ILD and SiD detector concept groups for conducting the full detector simulation studies. 
The speaker thankfully acknowledges the support of the Deutsche Forschungsgemeinschaft (DFG) through the Collaborative Research Center SFB 676, Particles, Strings and the Early Universe, subproject B1.

\end{document}